\documentclass[preprint,3p,times,twocolumn]{elsarticle}
\usepackage{epsfig}
\usepackage{amssymb}
\usepackage{amsmath}
\usepackage{dsfont}
\biboptions{sort&compress}

\usepackage{graphicx,hyperref}% Include figure files
\usepackage{bm}% bold math
\usepackage{subfigure}% bold math

\newcommand{\beq}{\begin{equation}}
\newcommand{\eeq}{\end{equation}}
\newcommand{\bea}{\begin{eqnarray}}
\newcommand{\eea}{\end{eqnarray}}
\newcommand{\nn}{\nonumber \\}

\newcommand\eqn[1]{(\ref{#1})}      % parentheses around the LaTex "ref" macro
\newcommand\Eqn[1]{Eq.~(\ref{#1})}  % includes ``Eq.'' in front
\newcommand\App[1]{\ref{#1}}

\newcommand{\C}{\mathcal{C}}
\newcommand{\Pn}{\mathcal{P}}

\newcommand{\bce}{\begin{center}}
\newcommand{\ece}{\end{center}}

\newcommand{\bl}{\,\bar{\!\lambda}}

\journal{Physics Letters B}

\begin{document}

\begin{frontmatter}



\title{Infrared dynamics in de Sitter space from Schwinger-Dyson equations}


\author{F. Gautier \corref{cor1}} \ead{fgautier@apc.univ-paris7.fr}
\author{J. Serreau \corref{cor1}} \ead{serreau@apc.univ-paris7.fr}%
\address{APC, AstroParticule et Cosmologie, Universit\'e Paris Diderot, CNRS/IN2P3, CEA/Irfu, Observatoire de Paris, Sorbonne Paris Cit\'e,\\ 10, rue Alice Domon et L\'eonie Duquet, 75205 Paris Cedex 13, France}

\begin{abstract}
We study the two-point correlator of an $O(N)$ scalar field with quartic self-coupling in de Sitter space. For light fields in units of the expansion rate, perturbation theory is plagued by large logarithmic terms for superhorizon momenta. We show that a proper treatment of  the infinite series of self-energy insertions through the Schwinger-Dyson equations resums these infrared logarithms into power laws. We provide an exact analytical solution of the Schwinger-Dyson equations for infrared momenta when the  self-energy is computed at two-loop order. Our findings encompass previously obtained results using either stochastic or Euclidean approaches. The obtained correlator exhibits a rich structure with a superposition of free-field-like power laws. We extract mass and field-strength renormalization factors from the asymptotic infrared behavior. The latter are nonperturbative in the coupling in the case of a vanishing tree-level mass.
 
 \end{abstract}


\end{frontmatter}
%\linenumbers

\section{Introduction}
\label{sec:intro}

Quantum field theory (QFT) in curved space-times is a subject of topical interest. Black hole physics and early Universe cosmology are paradigm examples where the laws of quantum mechanics and of gravity come into play, opening the possibility for intriguing new physical effects, such as the Hawking radiation \cite{Brout:1995rd} or the generation of primordial density fluctuations in inflationary physics \cite{Baumann:2009ds}. They also raise specific questions whose understanding brings a deeper insight into the fundamental laws at work.

The maximally symmetric de Sitter space has attracted a great deal of attention both because of its direct relevance for inflationary physics and because it already exhibits the specific features of curved geometries as compared to the flat Minkowski space-time, arising, e.g., from gravitational redshift and particle creation \cite{Bros:1994dn,Starobinsky:1994bd,Prokopec:2002jn,Onemli:2002hr,Brunier:2004sb,Boyanovsky:2005px,vanderMeulen:2007ah,Marolf:2010nz,Higuchi:2010xt}. Of particular interest is the case of light fields in units of the de Sitter radius, which have no analog in flat space and for which the limit of zero curvature is nonuniform due to nontrivial infrared effects. In this case, perturbative calculations of radiative corrections are plagued by large infrared/secular terms which call for resummation \cite{Starobinsky:1994bd,Tsamis:2005hd,Weinberg:2005qc,Polyakov:2009nq,Krotov:2010ma}. 

Various methods have been developed over the years to deal with similar issues in flat space, e.g., near a critical point \cite{Delamotte:2007pf}, for bosonic degrees of freedom at very high temperatures \cite{Blaizot:2003tw} or for systems out of equilibrium \cite{Berges:2003pc}. In recent years, a lot of activity has been devoted to adapt these techniques to nontrivial cosmological spaces, mainly for scalar fields. These include semiclassical stochastic methods \cite{Starobinsky:1994bd,Lazzari:2013boa}, renormalization group \cite{Burgess:2009bs,Kaya:2013bga}, two-particle-irreducible (2PI) \cite{Ramsey:1997qc,Riotto:2008mv,Tranberg:2008ae,Garbrecht:2011gu,Prokopec:2011ms,Parentani:2012tx}, or large-$N$ techniques \cite{Serreau:2011fu,Serreau:2013psa}, ladder-rainbow resummation \cite{Youssef:2013by}, or a field theoretic generalization of the Wigner-Weisskopf method \cite{Boyanovsky:2011xn,Boyanovsky:2012qs}. Useful information can also be gained by considering Euclidean de Sitter space \cite{Hu:1985uy,Higuchi:2010xt,Rajaraman:2010xd}. These studies reveal a realm of interesting infrared phenomena in de Sitter space.

A nontrivial problem in this context is to solve the Schwinger-Dyson equation for the two-point correlator for a given self-energy, that is to sum up the infinite series of self-energy insertions. This is a trivial step in the Minkowski vacuum---or, more generally in space- and time-translation invariant states---where the integro-differential Schwinger-Dyson equation is turned into a simple algebraic one in full $D$-dimensional momentum space, where $D=d+1$ is the space-time dimension. This is not so in de Sitter space despite its large degree of symmetry due to the noncommutativity of space- and time-translation generators. As a consequence, the mere inversion of the inverse propagator is a complicated problem. So far, attempts in this direction typically employ some ansatz solution in the deep infrared, see , e.g.,  \cite{Garbrecht:2011gu,Garbrecht:2013coa} (see also \cite{Jatkar:2011ju,Akhmedov:2011pj} for the case of massive fields).

In this Letter, we study this issue for the case of an $O(N)$ scalar field with quartic self-interactions. We consider the simplest nontrivial case with the (nonlocal) self-energy computed at two-loop order. Employing the physical momentum representation for de Sitter correlators (hereafter called the $p$-representation) \cite{Busch:2012ne,Parentani:2012tx} and the techniques developed in Ref. \cite{Serreau:2013psa}, we show that the Schwinger-Dyson equation, restricted to infrared momenta, is amenable to a one-dimensional integro-differential equation that can be exactly solved by analytical means. Using the analysis of Ref. \cite{Garbrecht:2013coa}, we check that our result reproduce those of either the stochastic or the Euclidean approaches for what concerns the field correlator in the coincidence limit in the case where perturbation theory is valid. Our results go beyond what has been previously obtained with these approaches and unravel the detailed structure of two-point correlators from horizon to deep superhorizon scales.

The Schwinger-Dyson equations resum the infinite series of perturbative infrared logarithms into modified power laws, in a way analogous to the generation of an anomalous dimension in critical phenomena.\footnote{A similar observation is made in \cite{Boyanovsky:2012qs}.} In de Sitter space, this can equivalently be seen as a mass correction \cite{Serreau:2013psa}. The resulting two-point correlator exhibits an interesting structure with a superposition of free-field-like power laws. The asymptotic infrared behavior is that of a noninteracting massive field in a Bunch-Davies vacuum state with renormalized mass and field strength. We compute the corresponding renormalization factors and show that they become nonperturbative in the coupling when the field is perturbatively light (massless) in units of the de Sitter radius. The two-loop approximation for the self-energy is not sufficient in that case and one should use nonperturbative approximation schemes, such as $1/N$-expansion or self-consistent 2PI approximations. We believe the tools and insight developed here are useful for such studies.

\section{General setting in the  p-rep\-re\-sen\-ta\-tion}
\label{sec:model}

We consider an $O(N)$ scalar field theory on the expanding Poincar\'e patch of de Sitter space in $D = d+1$ dimensions. In conformal time $-\infty < \eta < 0$ and comoving spatial coordinates $\bold{X}$, the invariant line element is given by (we set the Hubble scale $H=1$)
\beq
d s^2 = \eta^{-2}(-d \eta ^2 + d \bold{X} . d \bold{X}).
\label{eq:ds}
\eeq
The classical action is given by 
\beq
 \mathcal{S} = \int_x \left\{\frac{1}{2} \varphi_a \left(\square - m_{\rm dS}^2\right) \varphi_a - \frac{\lambda}{4!N}(\varphi_a\varphi_a)^2\right\},
 \label{eq:action}
 \eeq
with $\int_x = \int d^4 x \sqrt{-g}$ the invariant measure and where summation over repeated indices $a=1,\ldots,N$ is understood. Here, $\square$ is the appropriate Laplace-Beltrami operator and $m_{\mathrm{dS}}^2 = m^2 + \xi \mathcal{R}$ includes a possible coupling to the Ricci scalar  $\mathcal{R} = d(d+1)$. We consider a symmetric state such that $\langle\varphi_a\rangle=0$ and the correlator $G$ and the self-energy $\Sigma$ are diagonal, e.g., $G_{ab}=\delta_{ab}G$. In the rest of the Letter we assume a de Sitter invariant state.

The covariant inverse propagator is given by 
\beq
 \label{eq:first_2PI}
 G^{-1}(x,x') = G^{-1}_0(x,x') -  \Sigma(x,x'),
\eeq
where 
\beq
\label{eq:bareprop}
 iG_{0}^{-1}(x,x') = \left( \Box_x - m_{\mathrm{dS}}^2 \right) \delta^{(D)}(x,x'),
\eeq
with $\delta^{(D)}(x,x') = \delta^{(D)}{(x-x')}/\sqrt{-g(x)}$.
Extracting a possible local part from the self-energy\footnote{One may have to include more complicated structures in the local contribution to the self-energy when discussing ultraviolet renormalization. For instance in $D=4$, there appears a term $ \Box_x \delta^{(4)}(x,x^\prime)$ due to field-strength renormalization \cite{Brunier:2004sb}.}, one writes
\beq
\Sigma(x,x') = -i \sigma\delta^{(D)}(x,x') +\bar\Sigma(x,x'),
\label{eq:self_dec}
\eeq
where the local $\sigma$ part is constant for a de Sitter invariant state. We include it in a redefinition of the mass
\beq
 M^2 = m^2_{\mathrm{dS}} + \sigma \label{eq:gap}
\eeq 
and define, accordingly, the  propagator
\beq
\label{eq:massive}
 iG_{M}^{-1}(x,x') = \left( \Box_x - M^2 \right) \delta^{(D)}(x,x'),
\eeq
in terms of which we have
\beq
 \label{eq:first_2PI-M}
 G^{-1}(x,x') = G^{-1}_M(x,x') -  \bar\Sigma(x,x').
\eeq

Exploiting the spatial homogeneity and isotropy in comoving coordinates, one writes
 \beq
 G(x,x') = \int \frac{d^d K}{(2\pi)^d} \,\, e^{i\bold{K}\cdot({\bf X}-{\bf X}')} \tilde G(\eta , \eta^\prime , K).
\label{eq:fourier}
\eeq
De Sitter symmetries guarantee that the correlator admit the following $p$-representation \cite{Parentani:2012tx}
\beq
 \tilde G(\eta , \eta^\prime , K)=\frac{(\eta\eta')^{\frac{d-1}{2}}}{K}\hat G(p,p'),
\eeq
with $p = -K\eta$ and $p^\prime = - K \eta^\prime$ are the physical momenta associated to the comoving momentum $K$ at times $\eta$ and $\eta'$ respectively. Similarly, the $p$-representation of the self-energy is
 \beq
 \tilde \Sigma(\eta , \eta^\prime , K)=(\eta\eta')^{\frac{d+3}{2}}K^3\hat \Sigma(p,p').
\eeq

Solving the Schwinger-Dyson equation \eqn{eq:first_2PI-M} for the propagator $G$ for a given self-energy $\bar\Sigma$ in de Sitter space can be viewed as an initial value problem with initial data to be specified in the infinite past $\eta\to-\infty$. This can be formulated by introducing a closed contour in time---the so-called {\it in-in} formalism--- which allows one to conveniently grab together the various components of Green's functions \cite{Berges:2003pc}. Alternatively, in the $p$-representation, \Eqn{eq:first_2PI-M} becomes a flow equation in momentum, with initial data to be specified at $p\to+\infty$. Introducing a closed contour $\hat\C$ in momentum, the propagator reads \cite{Parentani:2012tx}
\beq
 \hat G(p,p')=\hat F(p,p')-\frac{i}{2}{\rm sign}_{\hat\C}(p-p')\,\hat\rho(p,p')
\eeq
where $\hat F$ and $\hat\rho$ denote the $p$-representations of the statistical and spectral two-point functions respectively. Here, the sign function is to be understood along the contour $\hat\C$. Notice the symmetry properties $\hat F(p,p')=\hat F(p',p)$, and $\hat \rho(p,p')=-\hat \rho(p',p)$. The self-energy $\hat\Sigma(p,p')$ admits a similar decomposition.

The Schwinger-Dyson equations read, in the $p$-representation \cite{Parentani:2012tx}
\bea
&&\hspace{-1.2cm}\left[\partial_p^2 + 1 - \frac{\nu^2 - 1/4}{p^2}\right] \hat F(p,p')+ \int_{p}^{\infty} \!\!\! d s \,  \hat \Sigma_\rho(p,s) \hat F(s,p')\nn
\label{eq:p_sd_F}
&&\hspace{2cm}=\int_{p'}^{\infty} \!\!\! d s \,  \hat \Sigma_F(p,s) \hat \rho(s,p')
\eea
and
\beq									
\left[\partial_p^2 + 1 - \frac{\nu^2 - 1/4}{p^2}\right] \hat \rho(p,p')	=\int^p_{p'} \! dq \,  \hat \Sigma_\rho(p,q)  \hat \rho(q,p') \label{eq:p_sd_rho},
\eeq
where we introduced (the last equality defines $\varepsilon$)
\beq
 \nu=\sqrt{\frac{d^2}{4}-M^2}\equiv\frac{d}{2}-\varepsilon.
\eeq
In the following, we consider light fields, i.e.,  $\varepsilon\approx M^2/d\ll1$.
The statistical function $\hat F$ encodes the information about the actual quantum state of the system. Having in mind an adiabatic switch-on of the interactions,\footnote{The authors of \cite{Polyakov:2009nq,Krotov:2010ma} have raised some objections against the possibility of adiabatically switching on the interactions in global de Sitter space. These do not apply to the case of the expanding Poincar\'e patch considered here.} the initial data corresponding to the Bunch-Davies vacuum are given by $\hat F(p,p^\prime) |_{p=p^\prime \to + \infty} =1/2$, $\partial_p  \hat F(p,p^\prime) |_{p=p^\prime \to + \infty} = 0$, 
$\partial_p \partial_{p^\prime} \hat F(p,p^\prime)|_{p=p^\prime \to + \infty}=1/2$. The nontrivial initial data for the spectral function is determined by the equal-time commutation relations: $\partial_p \hat  \rho(p,p^\prime)|_{p=p^\prime } = -1$. 

\section{The self-energy at two-loop}

Our purpose in this Letter is to invert the Schwinger-Dyson equation \eqn{eq:first_2PI-M} for the simplest nonlocal self-energy $\bar\Sigma$. We compute the latter in a loop expansion at two-loop order. Note that, by definition, local, tapole-like, diagrams, are all included in \eqn{eq:gap}. To cope with possible infrared divergences in the case $m_{\rm dS}^2\le0$, we may have to resum an infinite series of such tadpole diagrams. We leave the mass $M$ unspecified for the moment and perform a loop expansion for $\bar\Sigma$ in terms of the propagator $G_M$, \Eqn{eq:massive}.\footnote{Note that there can be no double counting issues since we are dealing with different topologies (tadpole vs. non-tadpole diagrams).}

\begin{figure}[t]
  \centering
  \includegraphics[width=.5\linewidth]{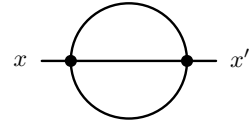}
  \caption{The nonlocal self-energy $\bar\Sigma(x,x')$ at two-loop order. The lines denote the propagator $G_M$, see \Eqn{eq:massive}.}
  \label{fig:self}
\end{figure}

The two-loop self-energy, represented in Fig. \ref{fig:self}, reads $\bar \Sigma =-\lambda^2 (N + 2)G_M^3/18N^2$. In the $p$-representation, one gets \cite{Parentani:2012tx}
\bea
&&\hspace{-1.4cm}\hat \Sigma (p,p') =  -\frac{ \lambda^2 (N + 2)}{18 N^2} \left(pp'\right) ^{{d-3}}\nn
\label{eq:pself}
&&\hspace{-.4cm}\times\int_{{\bf q},{\bf l}} \frac{ \hat G_M(qp,qp')\hat G_M(lp,lp')\hat G_M(rp,rp')}{qlr},
\eea
where $\int_{\bf q}\equiv\int d^dq/(2\pi)^d$ and $r = |\bold{e} + \bold{q} + {\bf l}|$, with $\bold{e}$ a unit vector. In the following we shall need the low momentum behavior of the statistical and spectral components of the self-energy \eqn{eq:pself}. Introducing an arbitrary scale $\mu\lesssim1$ to separate between super and subhorizon modes, one can show, using the method developed in Ref. \cite{Serreau:2013psa}, that the dominant contribution to the momentum integral in \Eqn{eq:pself} for $p,p'\lesssim\mu$ comes from momenta $qp,qp',lp,lp'\lesssim\mu$, which implies $qr,qr'\lesssim\mu$. One can thus evaluate the integral by using the leading infrared behavior of the propagator $\hat G_M$. Introducing the variables $x=\ln p/\mu$ and $x'=\ln p'/\mu$, the latter reads
\bea
\hat  F^{\mathrm{IR}}_{M}(p,p') &=&  \sqrt{pp'}\, \tilde F_\nu\,e^{-\nu(x+x')} \label{eq:f_loc_ir}, \\
 \hat \rho^{\mathrm{IR}}_{M}(p,p') &=& - \sqrt{pp'}\,\mathcal{P}_\nu(x-x')\label{eq:rho_loc_IR} ,
\eea 
where $\tilde F_\nu = \left[ 2^\nu \Gamma(\nu)\right]^2/{4\pi\mu^{2\nu}}$ and $\Pn_\nu(x) = \sinh(\nu x)/\nu$. The relevant momentum integrals are computed in the Appendix along the line of Ref. \cite{Serreau:2013psa}. We get\footnote{It is interesting to note the similar relations between statistical and spectral components of the various two-point function above. This may be a kind of fluctuation-dissipation relation.}
\bea
\label{eq:sigmaF}
 \hat \Sigma_F^{\rm IR}(p,p')&=&-(pp')^{- {3/2}} \,\tilde F_\nu\sigma_\rho\,s(x)s(x')\\
\label{eq:sigmarho}
 \hat \Sigma_\rho^{\rm IR}(p,p')&=&(pp')^{- {3/2}} \, \sigma_\rho\,\sigma(x-x')
\eea
with $s(x)=e^{-(\nu-2\varepsilon) x}$, $\sigma(x)={\cal P}_\nu(x) e^{-2\varepsilon|x|}$ and
\beq
\label{eq:sigma-rho-const}
 \sigma_\rho =\frac{\lambda^2}{\varepsilon^2} \frac{N+2}{24N^2} \frac{\Gamma^2(d/2)}{4\pi^{d+2}}.
\eeq
Here, we systematically neglect relative corrections of order $\varepsilon$ in the prefactors, but not in the $x$-dependence of the various functions. Indeed, such corrections become relevant at very large values of $|x|$ involved in convolution integrals. In particular, we emphasize the $e^{-2\epsilon|x|}$ term in \eqn{eq:sigmarho}, which is a nontrivial contribution from modes close to  the horizon in the loop integral \eqn{eq:pself} [see  \App{ap:two_loop}] and which plays a crucial role in obtaining the analytical solution below.

\section{The Schwinger-Dyson equations in the infrared}
\label{sec:schwinger}

We now proceed to solving equations \eqn{eq:p_sd_F} and \eqn{eq:p_sd_rho} for infrared momenta $p,p'\leqslant\mu$. Observe that, if the memory integral in \Eqn{eq:p_sd_rho} only involves infrared momenta, this is not so for \Eqn{eq:p_sd_F}. Here, we make the simplifying assumption that the dominant contributions come from momenta $p,p' \leqslant \mu$.\footnote{This is a widely used assumption in the literature. The rationale here is that one expects the high momentum contribution not to bring large infrared logarithms, which we are interested in. A detailed analysis of the high momentum contribution to a similar integral equation in the calculation of the four-point function in the large-$N$ limit has been performed in \cite{Serreau:2013psa}. Although the details are rather subtle, the naive expectation appears correct in that case and the high momentum modes only contribute a renormalization factor of order unity.} and we neglect the effect of the nonlocal self-energies for modes $p,p'\ge\mu$, resulting in the ``unperturbed'' solution $\hat F=\hat F_M$ for the latter. In practice, this amounts to switching on the nonlocal self-energy \eqn{eq:sigmaF} for $p,p'\le\mu$ and to match the high and low momentum solutions $\hat F_M$ and $\hat F$ at $p=p'=\mu$. 

Using standard manipulations \cite{Gautier:2012vh}, the general solution of \Eqn{eq:p_sd_F} can be written as 
\beq
\label{eq:Fsol}
 \hat F(p,p')=\int^\infty_p \!\!\!ds\!\int^\infty_{p'}\!\!\! ds' \,\hat \rho(p,s)\,\hat \Sigma_F^R(s,s')\,\hat \rho(s',p'),
\eeq
where, according to the discussion above,
\beq
 \Sigma_F^R(p,p')=\Sigma_F(p,p')\theta(\mu-p)\theta(\mu-p')-R_\mu(p,p').
\eeq
The second term on the right hand side ensures the correct boundary conditions at $p=p'=\mu$:
\bea
&&\hspace{-1.3cm} R_\mu(p,p')=A_\mu\,\delta(p\!-\!\mu)\,\delta(p'\!-\!\mu)+B_\mu\,\delta'(p\!-\!\mu)\,\delta'(p'\!-\!\mu)\nn
 &&\hspace{-.5cm} +C_\mu\left[\delta'(p\!-\!\mu)\,\delta(p'\!-\!\mu)+\delta(p\!-\!\mu)\,\delta'(p'\!-\!\mu)\right]
\eea
with $A_\mu=\partial_p \partial_{p^\prime} \hat F^{\rm IR}_M(p,p^\prime) |_{p=p^\prime =\mu}$, $B_\mu=\hat F^{\rm IR}_M(\mu,\mu)$ and $C_\mu=\partial_p  \hat F^{\rm IR}_M(p,\mu) |_{p=\mu} $. Using the explicit form \eqn{eq:f_loc_ir} of $\hat F^{\rm IR}_M$ one can show that, in terms of the variables $x$ and $x'$ introduced previously, 
\beq
 R_\mu(p,p')=(pp')^{-3/2}\tilde F_\nu\, r(x)r(x'),
\eeq
where
\beq
  r(x)=\delta'(x)-\nu\,\delta(x).
\eeq
Just as for $\hat \Sigma_F^{\rm IR}$ in \eqn{eq:sigmaF}, the factorization property of $R_\mu$ follows directly from that of $\hat F^{\rm IR}_M$.

The infrared dynamics reduces to a one-dimensional problem in terms of the variables $x$ and $x'$ introduced before. Indeed, one easily checks that the solutions of Eqs. \eqn{eq:p_sd_rho} and \eqn{eq:Fsol} take the form
\beq
\label{eq:statfull}
 \hat F^{\rm IR}(p,p')=\sqrt{pp'}\tilde F_\nu\left\{f_R(x)f_R(x')+\sigma_\rho\, f_\sigma(x) f_\sigma(x')\right\},
\eeq
with $f_R(0)=1$, $f_R'(0)=-\nu$, $f_\sigma(0)=f_\sigma'(0)=0$,
and 
\beq
 \hat \rho^{\rm IR}(p,p')=-\sqrt{pp'}\, \rho(x-x'),
\eeq
where the odd function $\rho(x)$ is such that $\rho'(0)=1$. It satisfies the following integro-differential equation
\beq
 \rho''(x)- \nu^2 \rho(x) =  \sigma_\rho\!\int_0^x \!\! dy\, \, \sigma(x-y)\,  \rho(y),\label{eq:rhonon_trivial}
\eeq
whereas $f_R(x)$ and $f_\sigma(x)$ in \eqn{eq:statfull} are given by ($x\le0$)
\beq
\label{eq:fR}
 f_R(x)=\int_x^0 \!\!dy\,\,  \rho(x-y)\,r(y)=\rho'(x)-\nu\,\rho(x)
\eeq
and
\beq
\label{eq:fbar}
  f_\sigma(x)=\int_x^0 \!\!dy\,\,  \rho(x-y)\,s(y).
\eeq
and are completely determined once $\rho(x)$ is known. We emphasize that Eqs. \eqn{eq:statfull}-\eqn{eq:fbar} are in fact quite general and hold whenever the self-energy assumes the form \eqn{eq:sigmaF} and \eqn{eq:sigmarho}.

To proceed, we write \Eqn{eq:rhonon_trivial} as an integral equation:
 \beq
\rho(x) = \rho_{M}(x) + \sigma_\rho\!\int_0^x \! dy \,  K(x-y) \,\rho(y),
\label{eq:rho_convol_x}
\eeq
where $\rho_M(x)={\cal P}_\nu(x)$ is the unperturbed solution, see \eqn{eq:rho_loc_IR}, and where we defined the kernel
\beq
  K(x) =  \int_0^x \!\!dy \,\, \rho_{M}(x-y) \,\sigma(y). \label{eq:kernel3}
\eeq
\Eqn{eq:rho_convol_x} resums the infinite series of self-energy insertions. It can be formally solved as 
\beq
	\rho(x) = \rho_{M}(x) +  \sigma_\rho\!\int_0^x \! dy \,   V(x-y)\,\rho_{M}(y) ,
	\label{eq:sol_eq}
\eeq
where the function $V$ satisfies the integral equation
  \beq
	 V(x)=  K(x) +  \sigma_\rho\!\int_0^x \! dy \,  K(x-y) \, V(y) .
	\label{eq:V_eq}
\eeq

\noindent The kernel \eqn{eq:kernel3} is readily obtained as
\beq
  K(x) = \frac{1}{4\epsilon\nu} \Big[ \Pn_{\nu_+} (x) - \Pn_{\nu_-}(x)  \Big]e^{-\varepsilon|x|}    ,  
 \label{eq:finalkernel}
 \eeq
where $\nu_\pm = \nu \pm \varepsilon$. The nontrivial task is to solve \Eqn{eq:V_eq}. Fortunately the exact solution can be found in closed form. It is straightforward to check that the desired solution is  
\beq
  V(x) = \frac{1}{4\varepsilon\tilde\nu} \Big[ \Pn_{\bar \nu_+} (x) - \Pn_{\bar \nu_- }(x)\Big] e^{-\varepsilon|x|},
\label{eq:vtilde}
\eeq
where 
\beq
\label{eq:nutilde}
\tilde\nu = \sqrt{\nu^2 + \frac{\sigma_\rho}{4\varepsilon^2}}
\eeq
and
\beq
\label{eq:nupm}
\bar \nu_{\pm}^2 =\nu^2\pm2\varepsilon  \tilde\nu+\varepsilon^2.
\eeq

The two-loop correction for the propagator is obtained by expanding in $\sigma_\rho$ at first nontrivial order. Such an expansion generates powers of $\sigma_\rho\ln x$ which may become large in the far infrared, invalidating the perturbative expansion. The proper treatment of the infinite series of self-energies insertions through the Schwinger-Dyson equations resums these large infrared logarithms in a well-defined expression \eqn{eq:vtilde}.

Having found $V(x)$, it is straightforward to obtain the solution \eqn{eq:sol_eq} of the Schwinger-Dyson equation \eqn{eq:rho_convol_x}. We get
 \bea
&&\hspace{-.6cm}\rho(x) = \frac{1}{2 \tilde \nu} \Bigg\{ ( \tilde \nu+\varepsilon)\Pn_{\bar \nu_+}(x) +(\tilde \nu-\varepsilon)\Pn_{\bar \nu_-}(x)  \nn 
\label{eq:rho_sol} &&\qquad+ \mathrm{sign}(x)\left[\Pn'_{\bar \nu_+}(x) -\Pn'_{\bar \nu_-}(x) \right]  \Bigg\}e^{-\varepsilon|x|} .
\eea
The exact expression of the statistical two-point function \eqn{eq:statfull} is then easily obtained from Eqs. \eqn{eq:fR} and\eqn{eq:fbar}:
\bea
\label{eq:F_solR}
  &&\hspace{-1.4cm}f_R(x)\!=\!\Bigg\{(\nu_-\!+\!\bar\nu_+)A_{\tilde\nu}(\bar\nu_+)e^{-\bar\nu_+x}\!+\!(\nu_-\!-\!\bar\nu_+)A_{\tilde\nu}(-\bar\nu_+)e^{\bar\nu_+x}\Bigg\}e^{\varepsilon x}\nn
  &&\hspace{-.6cm}+\,(\tilde\nu\to-\tilde\nu)
\eea
and
\beq
\label{eq:F_solsigma}
  f_\sigma(x)=\left\{\frac{A_{\tilde\nu}(\bar\nu_+)}{\nu_--\bar\nu_+ }e^{-\bar\nu_+x}+\frac{A_{\tilde\nu}(-\bar\nu_+)}{\nu_-+\bar\nu_+ }e^{\bar\nu_+x}\right\}e^{\varepsilon x}+(\tilde\nu\to-\tilde\nu),
\eeq
where we defined $A_{\tilde\nu}(z)=(z+\tilde\nu+\varepsilon)/4\tilde\nu z$. Notice that $\bar\nu_+\to\bar\nu_-$ when $\tilde\nu\to-\tilde\nu$.

\begin{figure}[t]
  \centering
  \includegraphics[width=.8\linewidth]{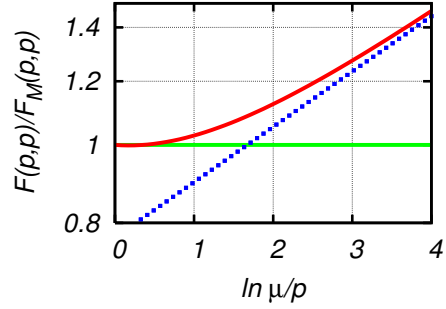}
  \caption{The resummed statistical function \eqn{eq:Fmassive} for equal momenta $p'=p\le\mu$, normalized by the free-field expression \eqn{eq:f_loc_ir}, on a logarithmic scale. It interpolates between the free-field behavior for $p\lesssim \mu$ and the asymptotic behavior \eqn{eq:f_loc_ir_eff} (dotted line) for deep infrared momenta $p\ll\mu$. The employed parameters are $d=3$ and $\varepsilon=\sigma_\rho=0.1$. }
  \label{fig:Fpp}
\end{figure}

The previous exact expressions take a simpler form already for moderate $|x|\gtrsim1$ and neglecting ${\cal O}(\varepsilon)$ relative corrections in numerical coefficients. In particular, we assume that $\varepsilon\tilde\nu\ll\nu^2$, which is always true in the range of applicability of the present calculation, as discussed below. For $|x|\gtrsim1$, we can write
$ \tilde\nu\,{\cal P}_\alpha(x)\pm{\rm sign}(x)\,{{\cal P}'_\alpha(x)}\approx\left(\tilde\nu\pm{\alpha}\right){\cal P}_\alpha(x),
$ and thus 
\beq
 \rho (x) = \Big\{c_+{\cal P}_{\bar\nu_+}(x)+c_-{\cal P}_{\bar\nu_-}(x)\Big\}e^{-\varepsilon |x|},\label{eq:rho_massive}
\eeq
with $c_\pm={(\tilde\nu\pm\nu)/2\tilde\nu}$. Similarly, we have ($x\le0$)
\bea
 f_R(x)&=&\left(c_+e^{-\bar\nu_+x}+c_-e^{-\bar\nu_-x}\right)e^{\varepsilon x},\\
 f_\sigma(x)&=&-\frac{1}{4\varepsilon\tilde\nu}\left(e^{-\bar\nu_+x}-e^{-\bar\nu_-x}\right)e^{\varepsilon x},
\eea
from which we deduce, using $c_+c_-=\sigma_\rho/16\varepsilon^2\tilde\nu^2$,
\beq
\label{eq:Fmassive}
 {\hat F^{\rm IR}(p,p')}\!=\!\!{\sqrt{pp'}\tilde F_\nu}\left\{c_+e^{-\bar\nu_+(x+x')}+c_-e^{-\bar\nu_-(x+x')}\right\}e^{\varepsilon (x+x')},
\eeq
with\footnote{Note that the dependence of each term on the right hand side of \eqn{eq:Fmassive} on the arbitrary scale $\mu\sim1$ is $\mu^{2(\bar\nu_\pm-\nu)}=1+{\cal O}(\varepsilon\ln\mu)$. A similar observation has been made in \cite{Serreau:2013psa,Boyanovsky:2011xn}.} $x=\ln p/\mu$ and $x'=\ln p'/\mu$.
Eqs. \eqn{eq:rho_massive} and \eqn{eq:Fmassive} are the central result of the present work. The resummation of nonlocal self-energy insertions produces a rich structure of the resummed propagator, as compared to the unperturbed one, Eqs. \eqn{eq:f_loc_ir} and\eqn{eq:rho_loc_IR}. Remarkably, it exhibits a superposition of unperturbed-like solutions with exponents $\bar\nu_\pm$, corrected by an $e^{-\varepsilon|x|}$ term which matters at large $|x|$. We see that both the spectral and statistical components of the propagator interpolate between an unperturbed (massive) behavior with exponent $\nu$ at moderate values of $|x|$ and an effective, unperturbed-like behavior with exponent $\bar\nu=\bar\nu_+-\varepsilon$ for very large $|x|$. This is illustrated in Fig. \ref{fig:Fpp}.

\section{Discussion}

To discuss the deep infrared behavior of the two-point correlators \eqn{eq:rho_massive} and \eqn{eq:Fmassive}, we consider the general local quadratic action
\beq
 {\cal S}_{\rm quad}= \frac{1}{2Z}\int_x \varphi_a \left(\square - \bar M^2\right) \varphi_a,
 \label{eq:actioneff}
 \eeq
where $\bar M$ is an effective mass and $Z$ a field-strength renormalization factor.\footnote{In Minkowski space, the definition \eqn{eq:actioneff} of the physical mass and field-strength renormalization coincide with the standard ones, namely the pole and the associated residue of the two-point correlator in $D$-dimensional momentum space. Equivalently, these parameters characterize the late-time behavior of the two-point correlators in real time. In de Sitter space, we do not have the luxury of a simple $D$-dimensional momentum representation and we thus adopt this alternative real time perspective. The physical mass $\bar M$ charaterizes the infrared, i.e. late time, power law behavior of the two-point correlators (more precisely the deviation from the free massless field exponent $d/2$) and the field-strength renormalization factor $Z$ measures the normalization of the correlators relative to the free-field case.} The corresponding correlators in a general de Sitter invariant state are easily obtained as, in the infrared limit,
\bea
 \hat  F_{\rm quad}^{\mathrm{IR}}(p,p')&=&  \sqrt{pp'}\, Z\,{\cal A}\, \tilde F_{\bar\nu}\,e^{-\bar\nu(x+x')} \label{eq:f_loc_ir_eff}, \\
 \hat \rho_{\rm quad}^{\mathrm{IR}}(p,p') &=& - \sqrt{pp'}\,Z\,\mathcal{P}_{\bar\nu}(x-x')\label{eq:rho_loc_IR_eff} ,
\eea 
with
\beq
\label{eq:effmass}
 \bar\nu=\sqrt{{d^2\over4}-{\bar M^2}}
\eeq
and where ${\cal A}$ quantifies the deviation from the Bunch-Davies vacuum (for which ${\cal A}=1$). It can be seen as a wavefunction renormalization. For the class of de Sitter invariant states, the so-called $\alpha$-vacua, ${\cal A}=|\cosh\alpha+e^{i\beta}\sinh\alpha|^2$ where $\alpha,\beta\in\mathbb{R}$. A deviation from unity can be seen as the result of particle production due to the field self-interactions, on top of that due to expansion. 

It is easy to check that, in the deep infrared regime $|x|\gg1$, the expressions \eqn{eq:rho_massive} and \eqn{eq:Fmassive} are of the form \eqn{eq:f_loc_ir_eff} and \eqn{eq:rho_loc_IR_eff} with, up to relative corrections of order $\varepsilon$,
\beq
\label{eq:IRproperties}
 {Z} =\frac{\tilde\nu+\nu}{2\tilde\nu}\,,\quad{\cal A}=1\quad{\rm and}\quad\frac{ \bar M^2}{M^2}=2-\frac{\tilde\nu}{\nu}.
\eeq 
Note that $Z\le1$ and $\bar M^2/M^2\le1$. Thus we find that the general solution of the previous section is well described, in the deep infrared, by the action \eqn{eq:actioneff}, which can thus be seen as the quadratic part of the effective action for deep infrared modes,\footnote{We emphasize that this only holds in the deep infrared regime. The general expressions \eqn{eq:rho_massive} and \eqn{eq:Fmassive} cannot be described by a local quadratic action.} and that there is no infrared renormalization of the vacuum state. 

The effective mass $\bar M$ describes the power law behavior of two-point correlators in the deep infrared or, equivalently their power law decay in space-time for asymptotically large separations. This differs from the so-called dynamical mass considered in previous works \cite{Starobinsky:1994bd,Rajaraman:2010xd,Garbrecht:2011gu}, which is defined as a measure of the field fluctuations at equal points:\footnote{More precisely, one considers the correlator $F(x,x')$ for de Sitter invariant distances $1\ll z(x,x')\ll1/\varepsilon$.}  $m_{\rm dyn}^2\propto 1/\langle\varphi^2(x)\rangle$. In particular, the latter receives contributions not only from the deep infrared regime \eqn{eq:f_loc_ir_eff}, but from all infrared modes and must be evaluated from the complete expression \eqn{eq:Fmassive}. We get
\begin{align}
 \frac{1}{N}\langle\varphi^2(x)\rangle&=F(x,x)=\int\frac{d^dp}{(2\pi)^d}\frac{\hat F(p,p)}{p}\nn
  &=\frac{\Gamma(d/2)}{2\pi^{{d/2}+1}}\mu^{2\varepsilon}\left(\frac{c_+}{d-2(\bar\nu_+\!-\!\varepsilon)}+\frac{c_-}{d-2(\bar\nu_-\!-\!\varepsilon)}\right)\nn
  &\equiv\frac{d\Gamma(d/2)}{4\pi^{{d/2}+1}m_{\rm dyn}^2},
\end{align}
where the second line corresponds to the contribution from infrared modes and the third line defines $m_{\rm dyn}$. Using \Eqn{eq:nupm} one easily checks that both terms in the second line contribute an infrared enhancement factor $1/\varepsilon$. We find, up to relative corrections of order $\varepsilon$,
\beq
\label{eq:mdyn}
 \frac{m_{\rm dyn}^2}{M^2}=\frac{4}{3}\left(1-\frac{\tilde\nu^2}{4\nu^2}\right)\le1.
\eeq
This clearly differs from the infrared mass $\bar M$, \Eqn{eq:IRproperties}. Instead we have $m_{\rm dyn}^2/\bar M^2=(2+\tilde\nu/\nu)/3\ge1$.
This shows that the dynamical mass $m_{\rm dyn}$ employed in the literature does not characterize the asymptotic power law behavior of two-point correlators in the infrared.\footnote{We point out that $m_{\rm dyn}$ is, in fact, a {\em static} quantity in that it measures the {\em amplitude} of the field fluctuations in the infrared. To our understanding, this is why it is directly accessible through Euclidean approaches. In contrast, $\bar M$ characterizes the momentum dependence of two-point correlators and is really, in that sense, of dynamical nature.}

We note that for coupling strong enough that \mbox{$\tilde\nu/\nu\ge2$}, one gets an unphysical solution with both $\bar M^2\le0$ and $m_{\rm dyn}^2\le0$. For instance, $F(x,x)\propto\int_{\bf p}\hat F(p,p)/p$ is infrared finite iff $\bar\nu<d/2\Leftrightarrow\bar M^2>0$. Similarly, the positivity of $\langle\varphi^2(x)\rangle$ implies that of $m_{\rm dyn}^2$. Using Eqs. \eqn{eq:nutilde} and \eqn{eq:IRproperties}, this implies $\sigma_\rho<3\varepsilon^2(d^2-\varepsilon^2)$.

\begin{figure}[t]
  \centering
  \includegraphics[width=.3\linewidth]{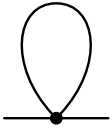}
  \caption{The self-consistent local self-energy $\sigma$. The line denotes the propagator $G_M$, see \Eqn{eq:massive}.}
  \label{fig:tadpole}
\end{figure}

To discuss the above results further, we now specify the mass parameter \eqn{eq:gap} which characterizes the propagator $\hat G_M$ entering the two-loop self-energy \eqn{eq:pself}. As mentioned previously, for $m_{\rm dS}^2\le0$ perturbation theory is ill-defined and one has to resum a mass. The simplest scheme to do so is to include the tadpole diagram of Fig. \ref{fig:tadpole} where the internal line corresponds to the propagator $G_M$. This is known as the Hartree approximation. In terms of the bare propagator \eqn{eq:bareprop}, it resums the infinite series of tadpole insertions---the so-called daisy and superdaisy diagrams---and leads to a self-consistent mass. In the small mass limit one has \cite{Hu:1985uy,Garbrecht:2011gu,Serreau:2011fu}
 \beq
M^2 = m^2_{\mathrm{dS}} +  \frac{\lambda(N+2)}{6N}\int_{\bf q}\frac{\hat F_M(q,q)}{q}\approx m^2_{\mathrm{dS}} +  \frac{c_d\lambda }{M^2} , \label{eq:massive_gap}
 \eeq
 where 
\beq
 c_d=\frac{N+2}{12N}\frac{d\,\Gamma(d/2)}{2\pi^{d/2+1}},
\eeq
in terms of which the parameter which controls the nonlocal self-energy corrections, see \Eqn{eq:nutilde}, reads
\beq
\label{eq:pertparam}
 \frac{\sigma_\rho}{4\varepsilon^2\nu^2}=\frac{6}{N+2}\Bigg(\frac{c_d\lambda}{M^4}\Bigg)^2 .
\eeq

There are two interesting limits to be considered. For a massive field with $c_d\lambda\ll m_{\rm dS}^4\ll1 $, perturbation theory is well defined. Introducing $\bl=c_d\lambda/m^4_{\rm dS}$, we get, at the order of interest,
\bea
{Z}&=& 1-\frac{3\bl^2}{2(N+2)}  \, \, ,  \\
\bar M^2&=&m_{\rm dS}^2\left(1+  \bl-\frac{N+5}{N+2}\bl^2\right)\,,\\
 m_{\rm dyn}^2&=&m_{\rm dS}^2\left(1+  \bl-\frac{N+4}{N+2}\bl^2\right).
\eea
The result for $m_{\rm dyn}^2$ agree with those from either the stochastic or Euclidean approaches in the perturbative regime for $N=1$, as recently shown in \cite{Garbrecht:2013coa}. This is a highly nontrivial test of our results since, as emphasized previously, the complete solution \eqn{eq:Fmassive} is needed to compute $m_{\rm dyn}$. We conclude that our findings equations \eqn{eq:rho_massive}-\eqn{eq:Fmassive} encompass the results of these approaches in the case where perturbation theory is applicable. 

The other interesting limit is that of massless or light fields, with $m_{\rm dS}^4\ll c_d\lambda$. In that case, one has $M^4\approx c_d\lambda$, that is $\sigma_\rho/4\varepsilon^2\nu^2\sim1$ and we get
\beq
\label{eq:masslesscase}
 \frac{\tilde\nu}{\nu}=\sqrt{\frac{N+8}{N+2}}.
\eeq
It follows that $Z$, $\bar M^2/M^2$ and $m_{\rm dyn}^2/M^2$ are all nonperturbative in $\lambda$.\footnote{Note that $\bar M^2>0\,\,\forall \,N$ and that $m_{\rm dyn}^2/M^2=N/(N+2)$.} These quantities actually measure the corrections to the local Hartree approximation from nonlocal self-energy insertions. The above result shows that the two-loop approximation for the latter is of the same order in coupling as the leading order (Hartree) result. This highlights the fact that the present perturbative evaluation of the self-energy \eqn{eq:pself} is not valid in that case. One should instead employ a nonperturbative approximation scheme, such as the $1/N$-expansion \cite{workinprogress} or self-consistent 2PI techniques \cite{Garbrecht:2011gu}.\footnote{The authors of \cite{Garbrecht:2011gu} employ a 2PI loop expansion at two-loop order in the case $N=1$, $d=3$ and implement a free-field ansatz in the infrared. Their result for the dynamical mass reads, in our notations, $m_{\rm dyn}^2/M^2=\sqrt 2$, very different from $m_{\rm dyn}^2/M^2=1/3$, obtained in the present perturbative approach. Although the comparison is meaningless owing to the nonperturbative nature of the light field case, we point out that the authors of \cite{Garbrecht:2011gu} did not consider the possibility of a field-strength renormalization in their ansatz. Furthermore, our results, Eqs. \eqn{eq:rho_massive} and \eqn{eq:Fmassive}, suggest that  a simple free-field ansatz might be questionable. This deserves further study.

In \cite{Boyanovsky:2012qs}, Boyanovsky considers the effect of the nonlocal two-loop self-energy in the case $N=1$, $d=3$ in the context of the Wigner-Weisskopf approach \cite{Boyanovsky:2011xn}. He concludes that the latter contributes a nontrivial decay width of single (quasi)particle states but does not give any mass correction. We believe that the apparent contradiction of this conclusion with the present findings as well as those of Refs. \cite{Garbrecht:2011gu,Rajaraman:2010xd} lies in the use of different definitions of mass. This requires further clarification.}

Finally, we mention that the case of negative tree-level mass square $m_{\rm dS}^2<0$, where $M^2=c_d\lambda/|m_{\rm dS}^2|$ \cite{Serreau:2011fu}, leads to an unphysical solution with $\bar M^2<0$. The two-loop approximation for the self-energy is certainly not valid in that case where the effective coupling $\sigma_\rho/2\varepsilon^2\nu^2\sim1/\lambda^{2}$ is nonperturbatively large.

\section{Conclusion}

We have obtained, for the first time, an analytical solution of the Schwinger-Dyson equations for the two-point correlators at superhorizon momenta in de Sitter space, in the case of the simplest (nonlocal) two-loop self-energy. This resums the infinite series of perturbative infrared/secular logarithms and results in a rich structure as compared to the noninteracting case, with modified infrared power laws. Our results encompass those of the stochastic and Euclidean approaches in the regime of validity of perturbation theory and reveals the detailed structure of the two-point statistical and spectral correlators from horizon to deep superhorizon momenta. It is remarkable that the asymptotic infrared behavior is that of a free massive field with renormalized mass and field strength. Although it has often been assumed to be the case in the literature, this had never been firmly established. In particular, the infrared field-strength renormalization has not been considered in previous studies. 

Our findings give new insight about the infrared structure of light scalar field correlators in de Sitter space. They provide useful guides for further investigations, e.g., using self-consistent resummation techniques \cite{workinprogress}. More generally, we believe the techniques developed here and in \cite{Serreau:2013psa} are useful tools for discussing infrared effects in the expanding de Sitter space. It is of interest to study the generalization of these tools to quasi de Sitter space relevant for inflationary cosmology. Finally, it would be interesting to extend the present techniques to the global de Sitter space and to study the resummation of the infrared divergences found in \cite{Polyakov:2009nq,Krotov:2010ma} and argued to signal the possible quantum instability of de Sitter space.

\section*{Acknowledgements}
We thank X. Busch, R. Parentani and A. Youssef for useful and interesting discussions.

\appendix{}
\section{The two-loop self-energy}
\label{ap:two_loop}

We write the two-loop self-energy of Fig. \ref{fig:self} as $\bar \Sigma=g\Pi G_M$, with $g=\lambda(N+2)/3N^2$ and where we introduced the one-loop function $\Pi=-\lambda G_M^2/6$. In the $p$-representation one has \cite{Parentani:2012tx}
\beq
\label{appeq:pself}
\hat \Sigma (p,p') =   g \left(pp'\right) ^{\frac{d-3}{2}}\!\!\int_{\bf q} \, \frac{r}{q}\,  \hat G_M(qp,qp')\hat \Pi(rp,rp') 
\eeq
and
\beq
\label{appeq:pi}
\hat \Pi (p,p') =  -\frac{\lambda}{6} \left(pp'\right) ^{\frac{d-3}{2}}\!\!\int_{\bf q} \, \frac{ \hat G_M(qp,qp')\hat G_M(rp,rp')}{qr},
\eeq
where $r=|{\bf q}+{\bf e}|$, with ${\bf e}$ a unit vector. The integral \eqn{appeq:pi} has been studied in detail in Ref. \cite{Serreau:2013psa}. In particular it has been shown there that the infrared behavior $p,p'\lesssim\mu$ is dominated by modes $qp,qp'\lesssim\mu$ under the integral. This condition obviously implies $rp,rp'\lesssim\mu$ and the integrand can thus be evaluated using the low momentum behavior of the propagator $G_M$, see Eqs. \eqn{eq:f_loc_ir} and \eqn{eq:rho_loc_IR}. The leading infrared behavior reads 
\bea
 \hat \Pi_F^{\mathrm{IR}}(p,p') &=& -\frac{\pi_\rho F_\nu}{(pp')^{\kappa+1/2}}  \label{eq:bubbleF} \, \, ,  \\
 \hat \Pi_\rho^{\mathrm{IR}}(p,p') &=& \frac{\pi_\rho}{\sqrt{pp'}}\,\Pn_{\nu}^\varepsilon\left(\ln\frac{p}{p'}\right)\label{eq:bubble_rho},
\eea 
where $\pi_\rho = \lambda F_\nu\Omega_d/6\varepsilon(2\pi)^d$ with $\Omega_d=2\pi^{d/2}/\Gamma(d/2)$,  $\kappa = \nu - \varepsilon$,  $F_\nu=\mu^{2\nu}\tilde F_\nu$ and $\Pn_\nu^\varepsilon(x)=\Pn_\nu(x)e^{-\varepsilon|x|}$. 

A similar line of reasoning can be applied to the integral \eqn{appeq:pself}. The dominant infrared behavior can be obtained by restricting the integral to the region $q\lesssim \min(\mu/p,\mu/p')$ and using the low momentum expressions \eqn{eq:f_loc_ir}, \eqn{eq:rho_loc_IR} and \eqn{eq:bubbleF} \eqn{eq:bubble_rho} for the integrand. The evaluation of the statistical component $\hat\Sigma_F$ involves the combination $\hat F_M^{\rm IR}\hat\Pi_F^{\rm IR}-\hat\rho_M^{\rm IR}\hat\Pi_\rho^{\rm IR}/4\approx \hat F_M^{\rm IR}\hat\Pi_F^{\rm IR}$ and one gets
\beq
\hat\Sigma_F^{IR}(p,p') = -\frac{g  F^2_\nu\pi_\rho}{ \left(pp'\right)^{\,\beta+{3/2}}} \int_{\bf q} \,\, \frac{1}{q^{2\nu}r^{2\kappa}} ,
\label{IRsigmaF_2}
\eeq
where $\beta=\nu-2\varepsilon$. The integral is rapidly convergent at high momentum and one can safely send the upper bound to infinity. The resulting integral can be evaluated by introducing Feynman parameters along the lines of \cite{Serreau:2013psa}. We get
\beq
 \int_{\bf q}\frac{1}{q^{2\nu} r^{2\kappa}}\!=\!\frac{\Omega_d}{2(2\pi)^d}\frac{\Gamma(\varepsilon)\Gamma(2\varepsilon)}{\Gamma(3\varepsilon)}\frac{\Gamma(\nu+\varepsilon)\Gamma(\beta)}{\Gamma(\nu)\Gamma(\beta+\varepsilon)}\!\approx\!\frac{3}{4\varepsilon}\frac{\Omega_d}{(2\pi)^d} \, \, , 
\eeq
where we neglected relative corrections of order $\varepsilon$. The final expression thus reads
\beq
\hat\Sigma_F^{IR}(p,p') = -\frac{\sigma_\rho  F_\nu}{ (pp^\prime)^{\,\beta+3/2}} \, \, , \label{oneloopsigmaF} 
\eeq
with $\sigma_\rho$ given in \Eqn{eq:sigma-rho-const}. This rewrites as \Eqn{eq:sigmaF}.

The spectral component $\hat \Sigma_\rho$ involves the combination $\hat F_M\hat\Pi_\rho+\hat\rho_M\hat\Pi_F$. It reads, in the case $p<p'$,
\bea
&&\hspace{-1.cm}\hat\Sigma_\rho^{IR}(p,p') =g\pi_\rho F_\nu(pp')^{\frac{d-3}{2}}\Pn_\nu\left(\ln\frac{p}{p'}\right)\nn
&&\hspace{-.5cm}\times\int_{|{\bf q}|<{\mu\over p'}}\left\{ \frac{1}{(pp')^\kappa} \frac{1}{r^{2\kappa}}
 + \frac{1}{(pp')^\nu}\left({p\over p'}\right)^\varepsilon \frac{1}{q^{2\nu}} \right\}.
\eea
The remaining integrals are easily performed (one can replace $r$ by $q$ in the denominator of the first term under the integral up to a correction of relative order $p'^2/\mu^2$). After some simple algebra and repeating the calculation for the case $p>p'$, we finally get
\beq
\hat\Sigma_\rho^{IR}(p,p') =  \frac{\sigma_\rho}{(pp')^{{3/2}}}\Pn_{\nu}^{2\varepsilon}\left(\ln\frac{p}{p'}\right),
\label{oneloopsigma}
\eeq
which is \Eqn{eq:sigmarho}.

\end{document}